# Termination-Preserved Ultra-high Tunneling Magnetoresistance in Altermagnetic $KV_2Se_2O$

*Junnan Guo,[1] Himanshu Mavani,[2] Wenhui Fang,[1] Jifeng Tang,[1] Wenhao Li,[1] Weikang Wu,[1*] Hui Li,[1]* Evgeny Y. Tsymbal,[2]* and Lishu Zhang[1]**

*[1]Key Laboratory for Liquid-Solid Structural Evolution and Processing of Materials, Ministry of Education, Shandong University, Jinan 250061, China*

*[2]Department of Physics and Astronomy & Nebraska Center for Materials and Nanoscience, University of Nebraska, Lincoln, Nebraska 68588-0299, USA*

E-mail: weikang_wu@sdu.edu.cn

E-mail: lihuilmy@hotmail.com

E-mail: tsymbal@unl.edu

E-mail: lishu.zhang@sdu.edu.cn

## Abstract

Altermagnets exhibit nonrelativistic spin splitting without net magnetization, establishing a new platform for next-generation spintronic devices. Although altermagnetic tunnel junctions (AMTJs) represent the most promising realizations, their practical applications are hindered by low tunnel magnetoresistance (TMR) ratios and strong sensitivity to interfacial configurations. Here, we systematically explore the transport properties and microscopic mechanisms of AMTJs based on the recently discovered *d*-wave altermagnet $KV_2Se_2O$. Using first-principles calculations and orbital-resolved analysis, we demonstrate that the synergy between compressed nodal-point like spin-degenerate channels and coplanar interfacial magnetic order yields an ultra-high intrinsic TMR above $10^5$% for all interfacial terminations. More importantly, K-termination effectively preserves bulk spin polarization through its unique passivation characteristics, leading to an ultra-high TMR up to $10^{12}$%. These results identify the coupling between momentum-space topology and interfacial passivation provides a reliable strategy for realizing giant magnetoresistive responses in altermagnetic spintronic devices.

# 1. Introduction

Magnetic tunnel junctions (MTJs) are the fundamental building blocks of magnetoresistive random access memory (MRAM), with each memory cell incorporating a discrete storage element.[1-3] Comprising two magnetic electrodes separated by a thin insulating barrier, an MTJ operates via the tunneling magnetoresistance (TMR) effect, in which the electrical resistance is governed by the relative magnetic configuration of the electrodes.[4, 5] Given that a high TMR ratio is indispensable for robust data readout, MTJs have traditionally employed ferromagnetic (FM) electrodes to provide substantial spin polarization, thereby generating large TMR ratios.[6, 7] However, the intrinsic stray magnetic fields associated with ferromagnetic order constrain device miniaturization and high-density integration.[8] While antiferromagnetic (AFM) materials present an appealing alternative due to their zero net magnetization and immunity to external magnetic perturbations, their fully compensated spin structures lead to vanishing net spin polarization, which makes it difficult to realize sizable TMR signals suitable for practical memory applications.[9-13]

The recent identification of altermagnetic (AM) materials, which exhibit non-relativistic spin splitting in their electronic band structure, provides a promising platform to break this deadlock.[14, 15] Unlike conventional collinear antiferromagnets, altermagnets break $PT$ or $U\tau$ symmetries (where $P$, $T$, $U$, and $\tau$ denote space inversion, time reversal, spin flip, and lattice translation, respectively), inducing momentum-dependent spin polarization even in the absence of spin-orbit coupling.[16-19] This characteristic grants altermagnets ferromagnet-like spin-polarization transport capabilities, while their antiparallel magnetic sublattices ensure antiferromagnet-like zero net magnetization. Consequently, even when the net tunneling current is spin-unpolarized, altermagnetic tunnel junctions (AMTJs) can still exhibit a giant TMR effect.[20] In AMTJs, the resistance state is determined by the relative orientation of the Néel vectors in the two altermagnetic electrodes, which can be configured as parallel (P) or antiparallel (AP).[21] The TMR effect arises from the matching and mismatching of the spin-polarized Fermi surfaces of the two electrodes within the two-dimensional Brillouin zone (2DBZ), and the TMR magnitude is directly governed by the overlap between these Fermi surfaces.[22, 23]

Currently, AMTJs based on $RuO_2$, MnTe, $NiF_2$, and CrSb have emerged, including all-altermagnetic architectures and altermagnetic barriers.[24-28] However, TMR ratios in these systems fail to meet industrial standards.[29-31] Furthermore, the transport process is often dominated by interfacial magnetic moment configurations rather than the intrinsic Néel vector of the altermagnet, leading to undesirable antiferromagnet-like behavior.[32-34] Most importantly, the microscopic physical origin of altermagnetic TMR, particularly the contribution mechanisms at the orbital level, lacks in-depth theoretical elucidation. To address these research gaps, the $d$-wave altermagnet $KV_2Se_2O$ is selected as the core subject of this study.[35] This material possesses a Néel temperature above room temperature and a unique quasi-two-dimensional flat Fermi surfaces that form only nodal-point like overlaps between spin channels, providing a natural structural foundation for achieving ultra-high TMR effects.[36] Moreover, its coplanar altermagnetic order greatly enhances transport stability and ensures robustness against interference from interfacial magnetic moment arrangements.

Despite these intrinsic advantages, the full TMR potential of $KV_2Se_2O$ remains largely underexplored. Existing studies mainly focus on bulk Fermi surface features.[37] Although the flat-band Fermi surface has been demonstrated to play an important role in achieving large TMR, the mechanisms responsible for sustaining spin polarization during transport have not yet been clearly elucidated. More importantly, a significant discrepancy exists between idealized AMTJ models and real physical systems.[38] As a $K^+$-intercalated material, $KV_2Se_2O$ tends to cleave along K or Se planes where interlayer bonding is weakest, whereas theoretical calculations adopt V–O atomic layers as the interface. Therefore, it is necessary to systematically investigate the influence of interfacial magnetic configurations and realistic cleavage surfaces on transport properties, while elucidating the underlying microscopic mechanisms from an orbital-resolved perspective[39-41]

In this study, we demonstrate that $KV_2Se_2O$-based AMTJs can achieve ultra-high TMR through the momentum-space selection mechanism and real-space interfacial magnetic moment matching. Using first-principles quantum transport calculations within a vacuum-barrier model, we elucidate the microscopic origin of this effect and quantify the role of interfacial terminations. Our findings demonstrate a universally ultra-high TMR exceeding $10^5$% for all interface terminations. Notably, the K-terminated interface achieves a remarkably high TMR of $10^{12}$% by preserving bulk spin polarization through its unique passivation effect. Extending the investigation to realistic heterostructures with $KTaO_3$ barrier yields TMR values in the range of $10^7$-$10^{12}$%, with the underlying transport mechanism remaining robust. These results highlight the critical roles of momentum-resolved spin channels and real-space interfacial configurations in enhancing TMR, and provide design principles for achieving extreme magnetoresistive responses in AMTJs.

# 2. Results and Discussion

### 2.1 Spin Channels and Design Principles

The TMR effect in MTJs originates from the symmetry and matching of spin-polarized states between the P and AP configurations. In AMTJs, where transport is governed by momentum-dependent spin splitting and alternating magnetic order, the TMR performance is primarily determined by the matching degree of momentum-space spin channels and the real-space arrangement of interfacial magnetic moments.

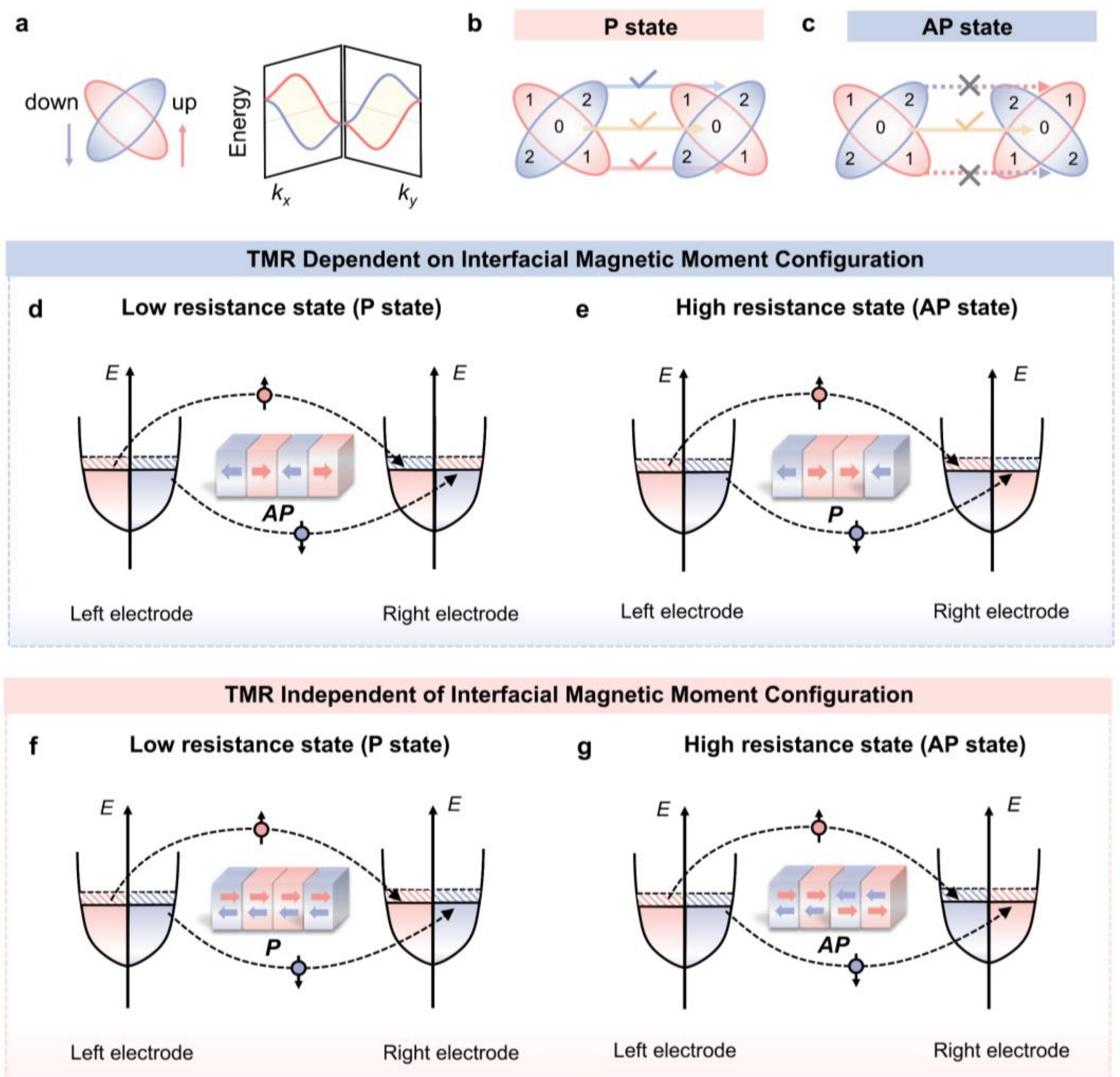

**Fig. 1. Spin channels and design principles. (a)** Schematic of the momentum-dependent spin distribution in a typical *d*-wave altermagnet. **(b, c)** Spin-resolved tunneling in altermagnet-based MTJs (AMTJs) under the ballistic transport framework. Regions 1 and 2 denote spin-up and spin-down channels; Region 0 represents the spin-degenerate overlap channel. **(d, e)** TMR dependent on interfacial magnetic moment configurations. **(f, g)** TMR independent of interfacial configurations in coplanar arrangements. The slashed regions represent the electronic states contributed by the interface. Pink and blue correspond to spin-up and spin-down electrons, respectively.

From the momentum-space perspective, the resistance originates from the matching and mismatching of the spin-polarized Fermi surfaces of the two electrodes projected onto the two-dimensional Brillouin zone (2DBZ) perpendicular to the transport direction. For a typical *d*-wave altermagnet, the two spin-resolved Fermi surfaces are related by a 90° rotation in momentum space (**Fig. 1**a). In the P configuration (**Fig. 1**b), the Néel vectors of the two electrodes are aligned, causing the spin-resolved Fermi surfaces to coincide, maximizing spin-matched transmission (**Regions 1** and **2**) and resulting in a low-resistance state. In contrast, in the AP configuration, reversal of one

Néel vector interchanges the spin characters, leading to momentum-space mismatch of the spin-resolved transmission channels and suppressed coherent tunneling, thus producing a high-resistance state (**Fig. 1**c). However, incomplete separation of opposite-spin Fermi surfaces inevitably leaves residual overlap regions (Region 0) in the projected 2DBZ.[42] These states remain transmissive even in the AP configuration and contribute nearly equally to the conductance of both P and AP states, thereby limiting the maximum achievable TMR. Reducing the area of this overlapping region is therefore expected to enhance the TMR. In the ideal ballistic limit, with no overlap between opposite spin-resolved transmission channels, the AP state transmission vanishes and TMR formally diverges.

Beyond Fermi surface topology constraints, the real-space arrangement of interfacial magnetic moments modulates the intrinsic bulk transport characteristics. In AMTJs based on $RuO_2$, $MnF_2$, and CrSb, atoms with opposite moments are often arranged non-coplanarly along the transport direction.[29, 30] Consequently, exposure of different magnetic atomic planes leads to a mismatch between interfacial moments and the electrode Néel vector. As illustrated in **Fig. 1**d, when the bulk is in a parallel configuration, the interface exhibits antiparallel coupling, suppressing the total electronic state in the P state. Conversely, interfacial parallel alignment during a bulk AP state enhances leakage (**Fig. 1**e). Such conflicts between the interface and bulk not only weaken the TMR effect but can also cause the AMTJs to degenerate into an interface-dominated antiferromagnetic-like junction, obscuring the intrinsic advantages of altermagnets.[34] However, if the magnetic atoms with opposite spins are localized within the same plane along the transport direction (**Figs. 1**f and g), the interfacial magnetic arrangement aligns with the electrode Néel vector. This decouples TMR from interfacial magnetic configuration interference, making it dependent solely on bulk altermagnetic features and specific atomic interfacial termination types.

Accordingly, achieving large TMR requires synergistic optimization by selecting altermagnets with minimized spin-degenerate regions on the Fermi surface and coplanar magnetic configurations. Meeting both criteria allows the AMTJ transport to bypass interfacial interference and recover its intrinsic bulk properties. This design principle establishes a clear strategy for optimizing the TMR effect and provides a clean physical model for investigating atomic-scale termination effects and orbital-related quantum transport mechanisms.

### 2.2 Electronic Structure of $KV_2Se_2O$

To validate the proposed design strategy, the *d*-wave altermagnet $KV_2Se_2O$ is selected as the electrode material, aiming to achieve a breakthrough in TMR performance by leveraging its unique nodal-point like spin-degenerate channels and coplanar alternating magnetic order.

The crystal structure of $KV_2Se_2O$ features a tetragonal layered lattice belonging to the $P4/mmm$

space group, as illustrated in **Fig. 2**a.[43] After geometric optimization, the calculated lattice parameters (a = b = 3.99 Å, c = 7.48 Å) are in excellent agreement with previous reports.[44] From an atomic-structural perspective, the Se atoms are located directly above and below the $V_2O$ planes, while adjacent $V_2Se_2O$ layers are separated by K atomic layers, giving rise to a well-defined layered stacking configuration. Magnetism originates from the V atoms within the $V_2O$ plane, where neighboring V atoms carry moments of equal magnitude but opposite direction, with the Néel vector oriented along the [001] axis. The two opposite-spin sublattices can be interconverted via $[C_2||C_{4z}]$ rotation or $[C_2||M_{1\bar{1}0}]$ mirror operations, forming a planar $d$-wave altermagnetic order. This arrangement ensures that sublattices with opposite spins reside entirely within the same atomic plane, forming a quintessential coplanar magnetic configuration that provides a structural foundation for avoiding interfacial magnetic mismatch.

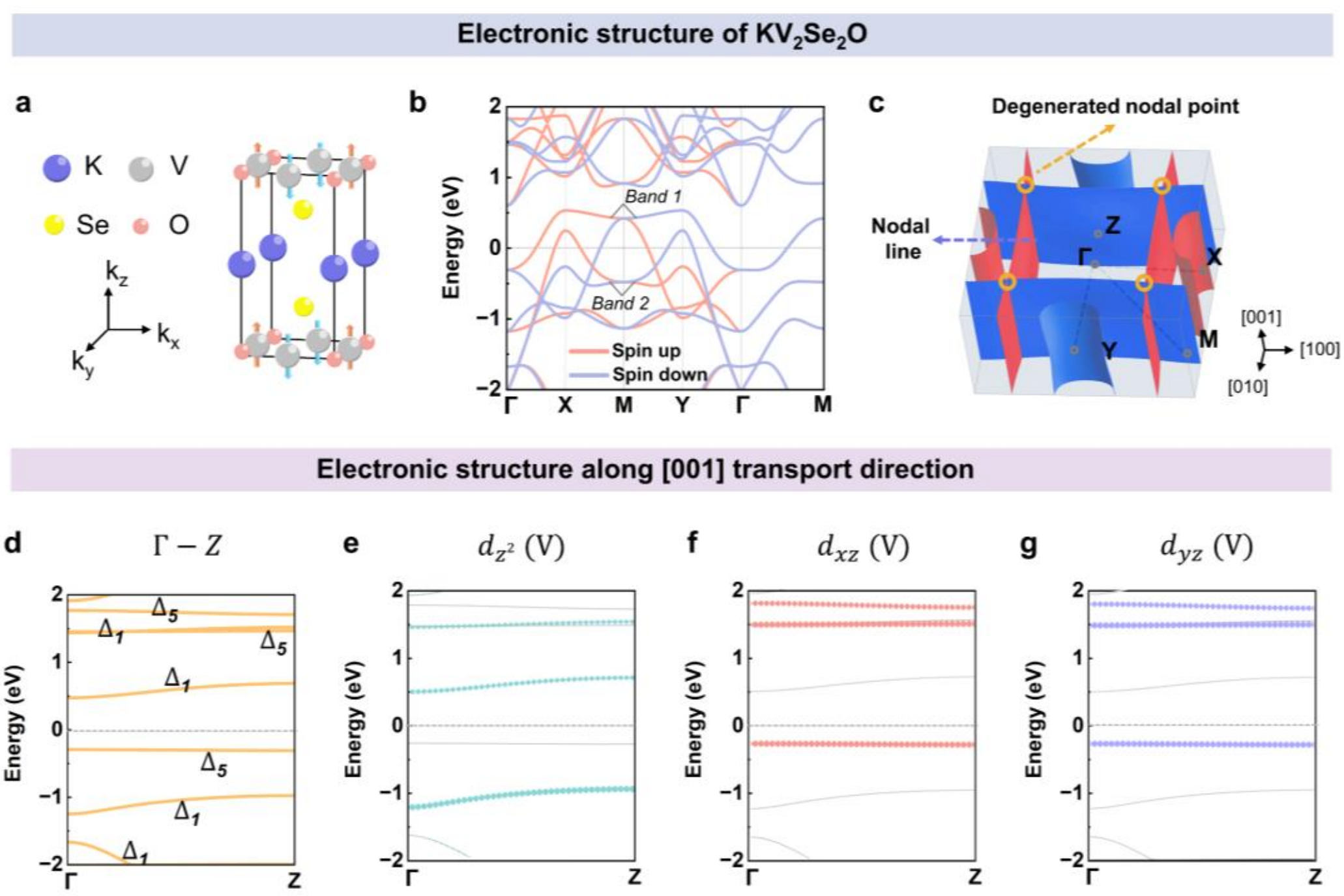

**Fig. 2. Electronic Structure of $KV_2Se_2O$**. **(a)** Crystal structure of $KV_2Se_2O$. Purple, grey, yellow, and pink spheres denote K, V, Se, and O atoms, respectively. The arrows on V atoms indicate the coplanar alternating magnetic order. **(b)** Calculated spin-resolved band structure of $KV_2Se_2O$. **(c)** Fermi surfaces of $KV_2Se_2O$. The overlap of spin-up and spin-down branches is localized at four isolated degenerate nodal points (marked by orange circles). **(d-g)** Symmetry-resolved band structure and orbital projections along the [001] transport direction (Γ–Z). **(d)** Symmetry-resolved band structure. **(e-g)** Orbital-projected bands for the V-$d_{z^2}$, V-$d_{xz}$, and V-$d_{yz}$ orbitals, respectively.

The calculated band structure clearly demonstrates the momentum-dependent spin splitting in $KV_2Se_2O$ (**Fig. 2**b). Along the high-symmetry paths Γ–X–M and Γ–Y–M, the energy bands exhibit opposite spin splittings, with a maximum splitting magnitude of approximately 1.6 eV. Two bands (Band 1 and Band 2) cross the Fermi level and contribute to the Fermi surface (**Fig. 2**c). Band 1 exhibits quasi-flat dispersions along the [001] direction, whereas Band 2 forms spin-polarized Fermi pockets near the X and Y valleys. The flatness of Band 1 plays a crucial role in yielding a large TMR effect, because it compresses the extended spin-degenerate regions into four isolated degenerate nodal points, thereby suppressing the leakage current in the AP state. As a result, when charge transport occurs parallel to these nodal lines (i.e., along [001]), the projected overlap between opposite spin-resolved transmission channels in the 2DBZ becomes effectively nodal-point like (**Figs. S4**a and **4**b). Thus, zero spin polarization occurs only at four isolated nodal points, while all remaining transmission channels are fully spin-polarized (**Fig. S4**c).

To further elucidate the microscopic origin of spin-dependent tunneling, the symmetry-resolved band structure of $KV_2Se_2O$ along the transport direction ([001]) is analyzed (**Figs. 2**d-g). Calculations reveal that transport channels near the Fermi level are mainly governed by the $\Delta_1$ and $\Delta_5$ irreducible representations, which correspond to symmetry-resolved states classified by the $C_{4v}$ point-group symmetry along the transport direction and associated with distinct orbital compositions (**Fig. 2**d). The $\Delta_1$ channel, primarily originating from the V-$d_{z^2}$ orbital (**Fig. 2**e), exhibits spin degeneracy along this high-symmetry path. Nevertheless, its contribution to the background current is negligible due to the aforementioned nodal-point like characteristics. Meanwhile, the $\Delta_5$ states exhibit a strong orbital symmetry constraint, which is essential for the altermagnetic effect. The spin-up channel is dominated by the $d_{xz}$ orbital of V atoms (**Fig. 2**f), while the spin-down channel is governed by the $d_{yz}$ orbital (**Fig. 2**g). Since the $d_{xz}$ and $d_{yz}$ orbitals are spatially orthogonal and have distinct transverse momentum distributions, such an orbital-dominated spin transport mechanism offers a strict symmetry-matching basis for employing $KV_2Se_2O$ as a high-performance AMTJ electrode.

### 2.3 Termination-Dependent TMR

To explore the intrinsic transport properties of $KV_2Se_2O$, we construct AMTJs with a 6 Å vacuum barrier (**Fig. 3a**). Three typical interfacial terminations, namely K-termination, Se-K-termination, and Se-termination, are considered due to their lower formation energy and experimental feasibility (**Fig. S3** and **Tab. S2**). Owing to the unique crystal structure of $KV_2Se_2O$, the interfacial magnetic moment orientation is independent of the exposed magnetic layer, which eliminates interfacial magnetic interference, enabling the transport properties to directly reflect the synergistic effects of bulk altermagnetic features and interfacial termination atomic types.

**Fig. 3**b illustrates the transverse momentum ($\boldsymbol{k}_{||}$)-resolved transmission for the K-termination AMTJ at the Fermi level. In both P and AP configurations, the transmission spectra for spin-up and spin-down electrons exhibit identical shapes with only a $\pi/2$ spatial rotation difference, directly

visualizing the *d*-wave symmetry inherent in $KV_2Se_2O$ bulk bands. In the P state, the transmission spectra present high-value regions with contours that highly coincide with the conduction channels. Conversely, in the AP state, overall transmission is suppressed due to the spin mismatch in momentum space, with the spectra shrinking into four isolated nodal-point like spots. This strong contrast in the tunneling landscapes leads to transmission in the P state markedly outweighing that in the AP configuration (**Figs. 3**c and d). Thus, the K-termination achieves an exceptionally large TMR ratio of $10^{12}$% (**Fig. 3**e).

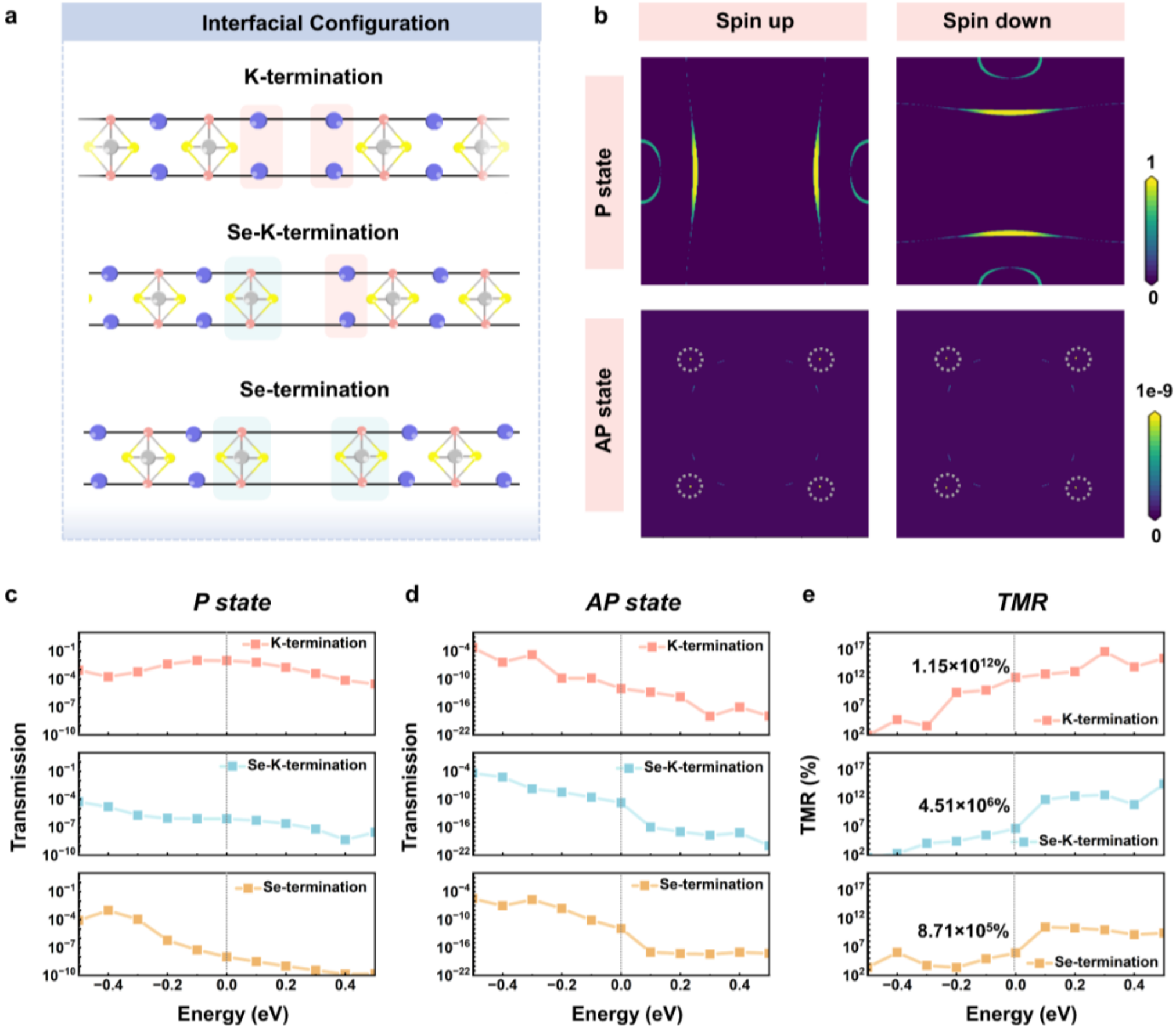

**Fig. 3. Termination-dependent tunnel magnetoresistance (TMR)**. **(a)** Interfacial configurations of three typical terminations: K-termination, Se-K-termination, and Se-termination. **(b)** Transverse momentum ($k_{\|}$)-resolved transmission for the K-termination configuration at the Fermi level. **(c, d)** Energy-dependent transmission spectra for the P state **(c)** and AP state **(d)** across the three termination types. **(e)** Calculated TMR ratios of the three terminations as a function of energy.

Other interfacial terminations render transmission probability extremely sensitivity to the atomic layer type. As shown in **Figs. 3**c-d, transitioning from K-terminated interface to Se-terminated one significantly reduces P state transmission and increases AP state leakage current. Consequently, the TMR of the Se-K-termination drops to $4.51\times10^{6}$%, while full Se-termination further plummets to the order of $10^{5}$% (**Fig. 3**e). This performance degradation is further corroborated by the local density of states (LDOS) distributions (**Fig. S5**), which confirm that the replacement of K-terminations triggers a drastic attenuation of the effective tunneling density of states within the barrier region. For all three terminations, TMR increases within the energy window, as the nodal-point like features shift toward the M point and eventually vanish (**Figs. S4**d-g). Nevertheless, despite large transmission fluctuations across terminations, $k_{||}$-resolved transmission spectra still retain the bulk Fermi surface feature (**Fig. S6**). Furthermore, the total current remains inherently spin-neutral regardless of the termination type (**Fig. S7**), which caused by the $d$-wave symmetry of the altermagnetic system. It is worth noting that the inclusion of spin-orbit coupling (SOC) does not alter the overall transport mechanism or the intrinsic origin of the ultrahigh TMR, as shown in **Fig. S9** and **Tab. S4**. Accordingly, results without SOC are mainly presented in the subsequent analysis to more clearly highlight the intrinsic transport mechanism of the system and avoid SOC-induced perturbations to the main physical behavior.

Ultra-high TMR realization depends not only on bulk Fermi surface structure but also crucially on interfacial atomic species. For comparison, transport properties along the [100] direction are also investigated (see **Figs. S10** and **S11**). Calculations show that while [100]-oriented AMTJs exhibit distinct spin splitting and certain robustness against interfacial variations, their TMR remains significantly lower than that along the [001] direction. This contrast highlights the application potential of the [001] orientation when coupled with specific interfacial terminations. Notably, it should be emphasized that even under the most unfavorable interfacial configurations, $KV_2Se_2O$-based AMTJs still outperform currently known AMTJ systems, fully demonstrating the great potential of this material in spintronic applications.

### 2.4 Termination-preserved TMR

Before analyzing the microscopic mechanisms underlying termination-dependent TMR, the thermodynamic stability of the considered surface terminations is first evaluated. Two surface slab models of $KV_2Se_2O$ corresponding to the K-termination and Se-termination are presented in **Fig. 4**a. The surface grand potential ($\Omega$) is employed to systematically assess the relative stability of these terminations as a function of the potassium chemical potential ($\Delta\mu_K$), with computational details provided in the Supporting Information (Section 3).[45] The resulting chemical potential phase diagram concerning $\Delta\mu_K$ and $\Delta\mu_{Se}$ is displayed in **Fig. 4**b. The calculations reveal that under K-rich conditions ($\Delta\mu_K$ > -0.889 eV), the K-termination exhibits the lowest surface grand potential ($\Omega_K$ = 0.0236), serving as the thermodynamic ground state of the system. In contrast, under Se-rich or K-poor conditions ($\Delta\mu_K$ < −0.889 eV), the Se-termination becomes the more stable configuration. However, the experimental synthesis of $KV_2Se_2O$ single crystals mainly relies on the KSe self-flux method, which provides a K-rich environment favorable for the formation of thermodynamically stable K-terminated interfaces.[35] Moreover, analysis of competing secondary phases including $K_2Se$, $V_2O_5$, and $VSe_2$ shows that these phases all lie outside the stability region of $KV_2Se_2O$, further verifying the experimental feasibility of realizing high-quality K-terminated interfaces.

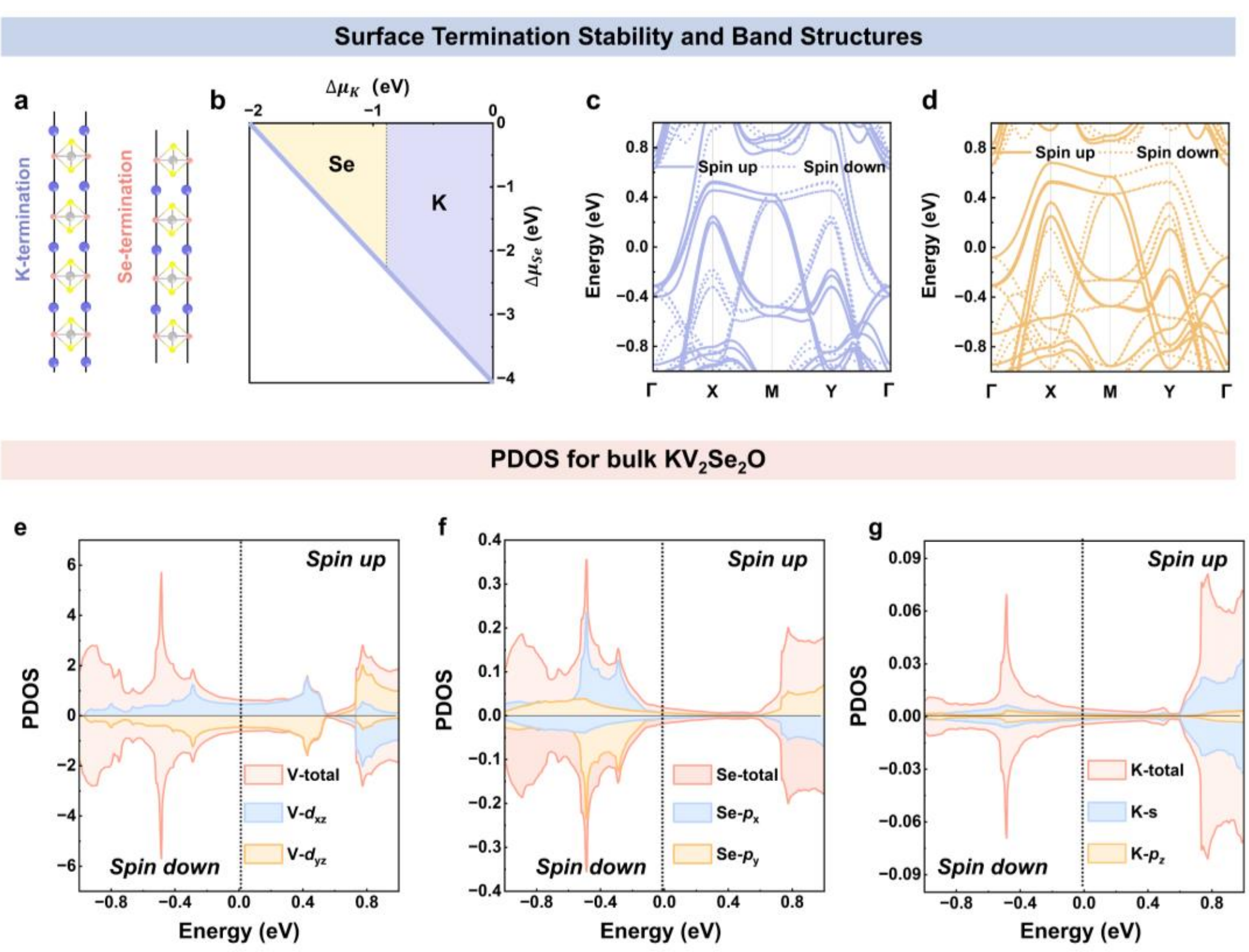

**Fig. 4. Microscopic physical mechanism of termination-dependent transport in $KV_2Se_2O$ AMTJs.** **(a)** Surface slab models of K-termination and Se-termination. **(b)** The chemical potential ($\Delta\mu_K$, $\Delta\mu_{Se}$) phase map. The shaded area is the thermodynamic stable range for the equilibrium growth of $KV_2Se_2O$, with yellow representing the stability region of the Se-terminated surface and purple representing that of the K-terminated surface. **(c, d)** Electronic band structures of K-termination **(c)** and Se-termination **(d)**. **(e-g)** Atom-resolved projected density of states (PDOS) for the spin-up and spin-down electronic states of V-element **(e)**, Se-element **(f)**, and K-element **(g)**.

Subsequently, electronic structures of different surface slab models are compared (**Figs. 4**c-d). Since the focus is on tunneling along the Z direction, the analysis primarily considers the band structure near the Γ point to assess transmission properties. The results show that K-terminated interface band features are highly consistent with bulk $KV_2Se_2O$, indicating that K-termination maximally preserves the bulk's intrinsic physical properties. In contrast, the Se-termination induces significant impurity levels within the band gap near the Fermi level, introducing deep-level defect characteristics to the system. The result is further validated by the comparison of the DOS in **Fig. S12**a, where K-termination does not alter DOS energy distributions, only slightly increasing intensity due to more atoms, while Se-termination generates substantial charge accumulation in the valence band region, directly corresponding to the defect levels in the band structure. Additionally, real-space electronic states confirm that the Se-terminated surface has numerous extra electronic states from dangling bonds, which significantly weaken the intrinsic altermagnetic properties (**Figs. S12**b and c).

Deep analysis of the orbital contributions from each atom in bulk $KV_2Se_2O$ further elucidates the microscopic mechanism, as depicted in **Figs. 4**e-g. In bulk $KV_2Se_2O$, altermagnetism primarily originates from the two orthogonal *d*-orbitals ($d_{xz}$ and $d_{yz}$) of the V atoms (**Fig. 4**e), followed by the *p*-orbital contribution from Se atoms (**Fig. 4**f), while K atoms' *s*-orbital contribution is extremely weak (**Fig. 4**g). This orbital weighting disparity determines the sensitivity of the electronic structure at the interface. Since Se atoms primarily contribute via *p*-orbitals, Se- termination exposure induces strong *p-d* orbital hybridization with the V atoms. This hybridization disrupts the local tetragonal symmetry of the V atoms and introduces incoherent scattering channels, acting as a bottleneck for TMR enhancement. This is further corroborated by the projected local density of states (PLDOS) along the transport direction (**Fig. S8**). With the substitution of Se atoms, the effective density of states penetrating the vacuum barrier increases significantly in the AP state, leading to a non-negligible leakage current. Conversely, in the P state, previously continuous conduction channels are disrupted by hybridization-induced scattering, substantially reducing tunneling state density. In contrast, K-termination exhibits unique electronic transparency. Since the DOS of K atoms mainly consists of *s* and $p_z$ orbitals (**Fig. 4**g), both extending along the *Z*-direction (the transport direction) and contributing minimally near the Fermi level, electrons of different spin orientations experience negligible transverse scattering when traversing the interface. In this near-ideal passivation environment, P-state conduction channels maintain continuous resonant tunneling features, while strict orbital orthogonality is preserved in the AP state, ultimately achieving an intrinsic TMR effect of up to $10^{12}\%$.

### 2.5 Enhanced TMR via Barrier Engineering

To investigate $KV_2Se_2O$ performance within realistic device architectures, AMTJs are constructed using $KTaO_3$ as the tunneling barrier (**Fig. 5**a). $KTaO_3$ is selected for its excellent lattice matching with $KV_2Se_2O$ (a mismatch of less than 1%) and wide bandgap characteristics.[46] Its band structure (Fig. 5b) exhibits no crossing bands at the Fermi level, confirming its superior insulating nature. Complex band structure calculations reveal that $KTaO_3$ also possesses $\Delta_1$ symmetry, which can enhance the selectivity of the tunneling process. (**Fig. S12**e). Moreover, $\boldsymbol{k}_{\parallel}$-resolved transmission calculations indicate that the rotational symmetry of $KTaO_3$ is highly compatible with that of $KV_2Se_2O$, and a high-transmittance window appears near the center of the Brillouin zone (**Fig. S12**f).[47] These features ensure effective lateral momentum conservation and the momentum-space polarization of the $KV_2Se_2O$ electrodes during transport, thereby enhancing the TMR effect.

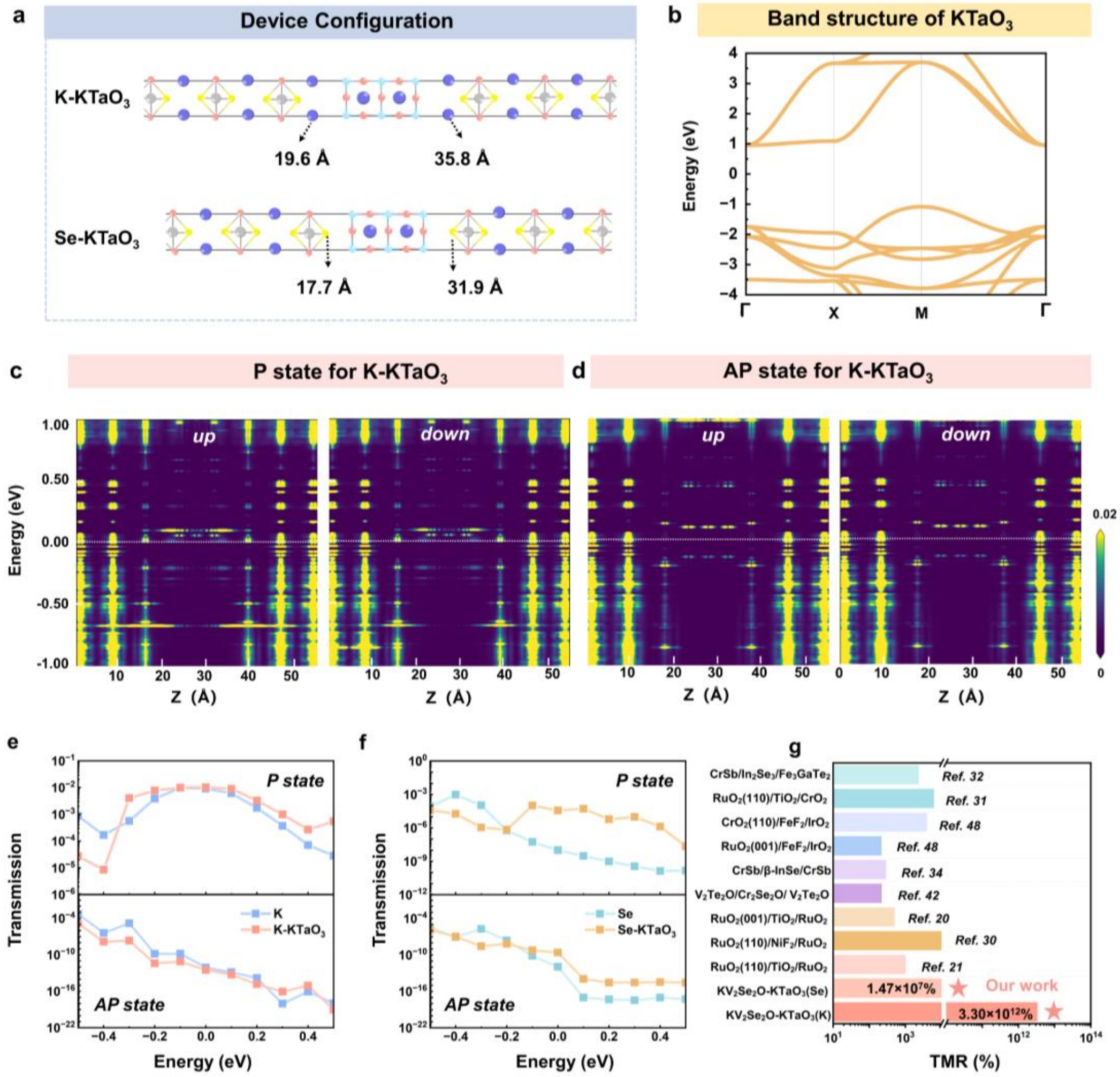

**Fig. 5. Enhanced TMR in $KV_2Se_2O/KTaO_3/KV_2Se_2O$ AMTJs**. (**a**) Device configurations of AMTJs with $KTaO_3$ barriers for K-termination and Se-termination. The outermost K and Se atoms at the interfaces are indicated in angstroms (Å). **(b)** Band Structure of $KTaO_3$. (**c, d**) Projected local density of states (PLDOS) for the K-terminated device in the P state **(c)** and AP state **(d)**. (**e, f**) Comparison of transmission spectra between vacuum-barrier and $KTaO_3$-barrier AMTJs for K-termination (**e**) and Se-termination (**f**). (**g**) Comparison of TMR ratios between this work and previously reported AMTJ systems.

The tunneling transport mechanism is intuitively reflected in the PLDOS of the device. As illustrated in **Figs. 5**b-c, taking the K-termination as an example, a clear energy gap persists at the Fermi level within the barrier region for P and AP states. This confirms that the $KTaO_3$ layer effectively decouples the Bloch states of the two electrodes, ensuring transport is dominated by coherent tunneling. The calculated layer-resolved DOS of the supercell also supports this mechanism (**Fig. S13**). In the P state, the barrier layer optimizes the interfacial matching (**Fig. 5**b), further enhancing the Fermi-level transmission coefficient compared to the vacuum-barrier case (**Fig. 5**d). Conversely, in the AP state, the absence of effective conduction channels between the left and right electrodes severely hinders electron propagation (**Fig. 5**c), dramatically reducing transmission (**Fig. 5**d). These dual factors substantially elevate the TMR value to $3.3 \times 10^{12}$%. Furthermore, the barrier layer also increases the TMR value of the Se-terminated AMTJ to $10^7$%. Although redundant surface states at the Se-interface increase tunneling probabilities for both P and AP states (**Figs. S14**a-b), the overall TMR remains high because the $KTaO_3$ barrier effectively preserves electron resonant tunneling characteristics in the P state (**Fig. 5**e). Additionally, *k*-resolved transmission coefficients show that the bulk $KV_2Se_2O$ Fermi surface features are precisely preserved after the introduction of the $KTaO_3$ barrier (**Figs. S14**c-d), with the total transmission remaining spin-neutral in both the P and AP states (**Fig. S15**). This further confirms that the giant TMR effect originates from the perfect synergy between the intrinsic physical properties of the electrodes and the high-quality interface.

To objectively assess the potential of $KV_2Se_2O$-based AMTJs, we compare them with recent representative theoretical studies on AMTJs (**Fig. 5**f).[20, 21, 30-32, 34, 42, 48] Existing AMTJs, whether through barrier optimization, half-metallic electrodes, or specific crystallographic orientations, TMR values typically below $10^5$%. However, the devices designed in this study achieve TMR values above $10^7$%, with the potential to leap to $10^{12}$% through rational interfacial termination design. Compared with recently reported AMTJs, particularly $KV_2Se_2O$-based AMTJs, this study elucidates the role of specific interface terminations in preserving spin polarization in altermagnetic systems. Specifically, the K-termination, owing to its unique passivation characteristics, maximally preserves the spatial orthogonality and intrinsic spin polarization of the V-3*d* orbitals in bulk $KV_2Se_2O$, thereby enabling exceptionally high TMR. Meanwhile, the intrinsically compressed nodal spin-degenerate channels, together with the coplanar interfacial magnetic order, provide the underlying physical basis for this termination-preserved behavior, allowing the ultra-high TMR to be fully realized. Overall, these results not only establish the theoretical performance upper limit of $KV_2Se_2O$-based AMTJs, but also provide clear design principles and guidelines for regulating altermagnetic tunneling transport via interface termination engineering.

### 2.6 Experimental Feasibility and Device Realization

To evaluate the experimental feasibility and device realization potential of $KV_2Se_2O$-based AMTJs, the robustness of the predicted ultra-high TMR is first examined under realistic conditions. Calculations show that the TMR remains stable in the range of $10^{10}$-$10^{12}$% despite variations in barrier thickness or the presence of interfacial defects, indicating excellent robustness against interfacial perturbations (**Figs. S16-17 and Tabs. S5-6**). Although an angular dependence is present, the TMR preserves its order of magnitude even under significant deviation from the ideal magnetic configuration, ensuring stable and reliable signal readout in practical operation (**Fig. S18 and Tab.**

**S7**). This angular response further provides a basis for the electrical detection of altermagnetic order and the development of high-sensitivity spintronic devices.

Experimentally accessible fabrication routes are available for $KV_2Se_2O$-based AMTJs. High-quality $KV_2Se_2O$ single crystals can be grown via the KSe self-flux method, from which atomically flat thin layers with a native K-terminated surface are obtained by mechanical exfoliation.[35] Alternatively, $KV_2Se_2O$ (001) thin films can be grown on lattice-matched substrates via molecular beam epitaxy (MBE) or pulsed laser deposition (PLD), where a K-rich growth environment stabilizes thermodynamically favorable K-terminated interfaces.[49, 50] A $KTaO_3$ tunneling barrier can then be deposited in situ, followed by epitaxial growth of the top $KV_2Se_2O$ electrode. Standard nanofabrication techniques, including ultraviolet lithography, $Ar^+$ ion etching, and Au/Pt electrode deposition, complete device fabrication and define the full heterostructure.[29]

For MRAM and related applications, the key advance of this system lies in field-free, fully electrical manipulation of the Néel vector.[51] The intrinsic altermagnetic spin-splitting effect (ASSE) in $KV_2Se_2O$ generates out-of-plane spin polarization, providing a physical foundation for spin-orbit torque (SOT)-driven switching.[52, 53] The Néel vector can be further tuned via current-induced effects, enabling reversible control of the spin-splitting torque (SST).[54] In addition, symmetry breaking of spin-group operations in altermagnetic systems gives rise to a magnetic orbital torque (MOT), which enhances the out-of-plane response and extends the capability of field-free control.[55] Through the spin Hall effect in heavy-metal/ $KV_2Se_2O$ heterostructures combined with interfacial symmetry breaking, current-induced Néel vector switching can be achieved. This mechanism has already been experimentally demonstrated in $RuO_2$-based AMTJs, further supporting its feasibility in altermagnetic devices.[29, 56]

Finally, to guide future experimental efforts, a general set of material screening criteria for AMTJs is extracted from the present framework. Ideal candidates should satisfy three essential requirements. A quasi-flat Fermi surface is required to confine spin-degenerate states in momentum space and enhance effective spin polarization. Coplanar interfacial magnetic order is required to eliminate Néel vector mismatch and stabilize transport. Chemically stable passivated terminations are required to preserve orbital selectivity and intrinsic spin polarization. Guided by these criteria, the tetragonal P4/mmm $XV_2Y_2O$ family (X = K, Rb, Cs; Y = S, Se, Te) serves as the most direct extension platform.[57-59] Among them, $RbV_2Te_2O$ and $Cs_{1-\delta}V_2Te_2O$ have been experimentally synthesized or theoretically predicted to host analogous altermagnetic order and quasi-flat band Fermi surfaces, showing great potential for excellent TMR responses.[60] Layered vanadium-based oxychalcogenides with identical V-O conducting motifs and tetragonal symmetry constitute a structurally well-defined candidate library, providing a unified theoretical framework for high-throughput screening and design of high-performance AMTJs.

## 3. Conclusion

In summary, this work reveals the ultra-high TMR in $KV_2Se_2O$-based AMTJs and its underlying microscopic mechanism. The quasi-flat dispersion along [001] compresses spin-degenerate regions into isolated nodal lines, minimizing opposite-spin channel overlap and suppressing transmission in the AP configuration. Meanwhile, the robust coplanar interfacial magnetic order ensures the alignment of interfacial magnetic moments with the bulk Néel vector, eliminating the interfacial magnetic mismatch. Through atomic-level interfacial engineering, K-termination yields an ultra-high TMR of $10^{12}$% by maximally preserving the spatial orthogonality and spin polarization of bulk V-*3d* orbital states. Even for Se-termination, where dangling bonds induce surface states, the system still exhibits a high TMR of $10^7$%, which is comparable to the highest values reported for altermagnetic systems. The performance enhancement upon introducing a $KTaO_3$ barrier further validates the robust transport stability of the junction under realistic device conditions. This study provides theoretical guidance and effective regulation strategies for the design of high-performance altermagnetic spintronic devices.

## 4. Methods

The geometric optimization, electronic structure calculations, and quantum transport simulations of $KV_2Se_2O$ are performed using the QuantumATK package.[61, 62] The exchange–correlation interaction is described within the spin-polarized generalized gradient approximation (SGGA) using the Perdew–Burke–Ernzerhof (PBE) functional.[63] All atoms are described using a High basis set together with PseudoDojo pseudopotentials.

For geometry optimization and electronic structure calculations, the Brillouin zone is sampled using a dense 25 × 25 × 15 Monkhorst–Pack *k*-point mesh. The real-space grid cutoff for the charge density is set to 80 Hartree, and the electronic temperature is fixed at 300 K. Structural relaxation is continued until the residual force on each atom is smaller than 0.01 eV/Å. With these parameters, both the optimized geometries and the resulting band structures show good agreement with available experimental data, ensuring the reliability of the calculations. The effects of Hubbard-U and SOC are further tested, and both are found to have negligible influence on the band structure (**Fig. S**1).

Quantum transport properties are investigated within the density functional theory (DFT) combined with the nonequilibrium Green's function (NEGF) formalism.[64-66] In the Green's function approach, an infinitesimal imaginary broadening parameter of 0.001 *Hartree* is introduced to ensure numerical stability of the energy integration and to avoid unphysical oscillations. The device is modeled as a two-probe system, consisting of a central scattering region sandwiched between two semi-infinite symmetric electrodes. For transport calculations, the real-space grid cutoff is increased to 150 Hartree. In the self-consistent transport calculations, a 15 × 15 × 150 $k$-point mesh is used. For the calculation of $\boldsymbol{k}_{\parallel}$-resolved transmission, the two-dimensional Brillouin zone is sampled using an dense meshes of 500 × 500 $k$ points, which has been verified by convergence tests (**Fig. S**2 and **Tab. S**1).

In crystalline MTJs operating in the ballistic regime and in the absence of spin-orbit coupling (SOC), both spin and the transverse wave vector $\boldsymbol{k}_{\parallel}$ is conserved. Consequently, the longitudinal tunneling conductance per unit area for spin $\sigma$ is given by $g^{\sigma} = \frac{e^2}{h}\sum_{\boldsymbol{k}_{\parallel}} N_{\parallel}^{\sigma}(\boldsymbol{k}_{\parallel})$. Here $N_{\parallel}^{\sigma}(\boldsymbol{k}_{\parallel})$ represents the spin-resolved number of conduction channels at transverse wave vector $\boldsymbol{k}_{||}$. We then define the $\boldsymbol{k}_{\parallel}$-resolved spin polarization $p_{\parallel}(\boldsymbol{k}_{\parallel})$ as:

$$p_{\parallel}(\boldsymbol{k}_{\parallel}) = \frac{N_{\parallel}^{\uparrow}(\boldsymbol{k}_{\parallel}) - N_{\parallel}^{\downarrow}(\boldsymbol{k}_{\parallel})}{N_{\parallel}^{\uparrow}(\boldsymbol{k}_{\parallel}) + N_{\parallel}^{\downarrow}(\boldsymbol{k}_{\parallel})}.$$

The TMR effect in MTJs fundamentally arises from the $\boldsymbol{k}_{\parallel}$ resolved spin polarization $p_{\parallel}(\boldsymbol{k}_{\parallel})$ matching conditions of two metal electrodes in the P and AP configurations.

TMR is defined as:

$$TMR = \frac{T_P - T_{AP}}{T_{AP}} \times 100\ \%$$

where $T_P$ and $T_{AP}$ denote the total transmission in the P state and AP state, respectively.

## Supporting Information

Supporting Information is available.

## Acknowledgments

We would like to acknowledge the support from the National Natural Science Foundation of China (NNSFC) (Grant Nos. 52501308), the Key Research and Development Plan of Shandong Province (Grant Nos. 2025CXGC020107 and 2021SFGC1001), the Key Technology Research and Development Program of Shandong Province (No. 2025CXGX010406) and the Natural Science Foundation of Shandong Province (No. ZR2025QC1106). This work is also supported by the Special Funding in the Project of the Taishan Scholar Construction Engineering and the program of Jinan Science and Technology Bureau (2020GXRC019) as well as new material demonstration platform construction project from Ministry of Industry and Information Technology (2020-370104-34-03-043952-01-11). Research at University of Nebraska-Lincoln (UNL) was supported by the National Science Foundation (Grant No. DMR-2316665) and UNL's Grand Challenges catalyst award.

## Conflict of Interest

The authors declare no competing financial interest.

## Data Availability Statement

The data that support the findings of this study are available from the corresponding author upon reasonable request.